\documentclass{aastex6}
\usepackage{amsfonts,amsmath,amstext,graphicx,subfigure,color,bm}
\usepackage[]{hyperref}

\begin{document}

\title{Fast Radio Bursts from the collapse of Strange Star Crusts}

\author{Yue Zhang$^{1}$}
\author{Jin-Jun Geng$^{1,2}$}
\author{Yong-Feng Huang$^{1,2}$}
\affil{
	$^1$School of Astronomy and Space Science, Nanjing University, Nanjing 210023, China\\
	$^2$Key Laboratory of Modern Astronomy and Astrophysics (Nanjing University), Ministry of Education, China
	}

\begin{abstract}
Fast radio bursts (FRBs) are transient radio sources at cosmological distances.
No counterparts in other bands have been observed for { non-repeating FRBs}.
Here we suggest the collapse of strange star crusts as a possible origin for FRBs.
Strange stars, which are composed of almost equal numbers of u, d, and s quarks,
may be encapsulated by a thin crust of normal hadronic matter. When a strange star
accretes matter from its environment, the crust becomes heavier and heavier.
It may finally collapse, leading to the release of a large amount of magnetic energy
and plenty of electron/positron pairs on a very short timescale.
Electron/positron pairs in the polar cap region of the strange star can be accelerated
to relativistic velocities, streaming along the magnetic field lines to form a thin shell.
FRBs are produced by coherent emission from these electrons when the shell is expanding.
Basic characteristics of observed FRBs can be explained in our model.
\end{abstract}


\section{Introduction} \label{1}
Fast radio bursts (FRBs) are a new kind of phenomena that were discovered in the past decade \cite{2007Sci...318..777L, 2013Sci...341...53T, 2012MNRAS.425L..71K}.
These transient bursts have flux densities of $S_{\rm{\nu}} \sim$ a few Jy at frequencies of $\nu_{\rm{FRB}} \sim 1\ \rm{GHz}$,
with the waveband width of $\Delta \nu_{\rm{FRB}} \sim$ several hundred $ \rm{MHz} $.
Their durations $\delta t $ are typically a few $\rm{ms}$, indicating a rather compact region of emission.
The observed high dispersion measurements (DMs) of $\sim 500 - 1000 \ \rm{pc\ cm^{-3}} $ are well above
the contribution from our Galaxy for several FRBs detected at
high-galactic-latitude ($\geq 40^{\circ}$) \cite{2002astro.ph..7156C, 2014ApJ...785L..26L},
suggesting that the sources are at cosmological distances of $d \sim \rm{Gpc}$ with redshifts of $z \sim 0.5 - 1$.
Hence the isotropic luminosities in radio waves ($L_{\rm{FRB}}$) are estimated
as $\sim 10^{42}$ --- $10^{43} \ \rm{erg \ s^{-1}}$, with the total isotropic energy released in a
typical burst being $E_{\rm{FRB}} \sim 10^{39}$ --- $10^{40}\ \rm{erg}$.
The event rate is estimated to be $\sim 2 \times 10^{3}\ \rm{sky^{-1} \ day^{-1}}$ \cite{2018MNRAS.475.1427B}.
The brightness temperatures of FRB sources can be as high as $T_{B} \ge 10^{36}\ \Gamma^{-2} \ \rm {K}$ \cite{2014ApJ...785L..26L, 2014PhRvD..89j3009K},
where $\Gamma$ is the Lorentz factor of the emitting material.
Such an extremely high temperature is far above the Compton limit for incoherent synchrotron radiation,
thus a coherent origin shall be considered \cite{2016PhRvD..93b3001R}.
On the other hand, no counterparts in other wavebands have been detected to associated with { non-repeating} FRBs hitherto.

The origin of FRBs still remains unclear, but a number of models trying to interpret these
enigma phenomena have been proposed, e.g., magnetar giant flares \cite{2014ApJ...797...70K},
the collapses of magnetized supramassive rotating neutron stars \cite{2014A&A...562A.137F, 2014ApJ...780L..21Z},
binary neutron star mergers \cite{2013PASJ...65L..12T}, binary white dwarf mergers \cite{2013ApJ...776L..39K}, collisions between neutron stars and asteroids/comets \cite{2015ApJ...809...24G}, collisions between neutron stars and white dwarfs \cite{ 2017arXiv171203509L}, and evaporation of primordial black holes \cite{2014PhRvD..90l7503B}.
Some of these models are catastrophic and the original central engines are destroyed completely by the bursts.
The discovery of the repeating FRB source, i.e., $\rm{FRB} \ 121102 $ \cite{2016Natur.531..202S, 2016ApJ...833..177S},
presents new interesting clues to FRBs.
Several elaborate models have been put forward to explain
its repeating behaviors, e.g., highly magnetized pulsars traveling through asteroid belts \cite{2016ApJ...829...27D},
neutron star-white dwarf binary mass transfer \cite{2016ApJ...823L..28G}, and star quakes of pulsars \cite{2018ApJ...852..140W}.
However, it is wondered whether $\rm{FRB} \ 121102 $ is representative of FRBs since its features are
different from others \cite{2018ApJ...854L..12P, 2018Natur.553..182M}.
Naturally, people speculate that repeating FRBs such as $\rm{FRB} \ 121102 $ may be a separate kind of FRB sources.

Here we propose a new model for FRBs. We argue that the collapse of strange star (SS)
crusts can also explain the main features of FRBs. The structure of this paper is as follows.
In Section \ref{2}, the process of SS crust collapse is illustrated.
The emission mechanism that leads to observable FRBs is described in Section \ref{3}.
Possible counterparts in other wavebands are discussed in Section \ref{4}.
Finally, Section \ref{5} is a brief summary and discussion.

\section{Collapse Process of SS Crust}  \label{2}

It has been conjectured that strange quark matter (SQM), a kind of dense material composed of
approximately equal numbers of up, down, and strange quarks, may have a lower energy per baryon
than ordinary nuclear matter (such as $^{56} \rm{Fe}$) so that it may be the true ground state
of hadronic matter \cite{1984PhRvD..30..272W, 1984PhRvD..30....1A}.
If this hypothesis is correct, then neutron stars (NSs) may actually be ``strange stars'' \cite{1986ApJ...310..261A}.
The bulk properties of strange stars and neutron stars are rather similar in the typical mass range
of $1 <= M / M_{\odot} <= 1.8$, and it is very difficult to discriminate between them \cite{1986A&A...160..121H}.
Although several methods have been proposed to distinguish strange stars from neutron
stars \cite{ 2015ApJ...804...21G, 2018RAA....18...24L}, no definitive conclusions have been drawn yet.

At the strange star SQM surface, the density reduces from $\sim 5 \times 10^{14}\ \rm{g\ cm^{-3}}$ to zero abruptly.
The thickness of the SQM surface is of order $1 \ \rm{fm}$ due to strong interaction between quarks,
while electrons can stretch up to several hundred $\rm{fm}$ beyond the surface since they are
bounded electromagnetically. An extremely intense electric field ($\sim 5 \times 10^{17}\ \rm{V/cm}$)
is induced by charge separation near the SQM surface \cite{1986ApJ...310..261A}.
The outward-directed electric field can polarize a layer of nearby normal matter and
provide a force overwhelming the gravity \cite{1997A&A...325..189H,  2005PhRvD..72l3005S}.
As a result, a thin crust composed of normal hadronic matter may exist and obscure the whole
surface of the SQM core. It has been shown by \cite{1997A&A...325..189H} that the maximum density
at the bottom of the crust should be significantly less than the so called neutron drip density.
For a typical SS with a radius of  $r \sim 10\ \rm{km}$, a mass of $M \sim 1.4\ M_{\odot}$ and a
surface temperature of  $T_{\rm{S}} \sim 3 \times 10^{7} \ \rm{K}$, the crust mass is usually
in the range of $M_{\rm{c}} \sim 10^{-7}\ M_{\odot}$ --- $10^{-5}\ M_{\odot}$ and its thickness is
about $l \sim 2 \times 10^{4}\ \rm{cm}$.

The distance between the bottom of the crust and the surface of the SQM core shall be at
least $\sim 200\ \rm{fm}$ so that the rate at which ions penetrate the gap through the
tunneling effect is low enough to ensure the stabilization of the crust \cite{1986ApJ...310..261A}.
If the mass of the crust increases continuously via some accreting process, then the gap between the
crust bottom and the SQM surface will become narrower and narrower to counterpoise the gravity of the
crust \cite{1997A&A...325..189H}. Once the gap is less than $\sim 200\ \rm{fm}$, a large portion of
ions could penetrate the Coulomb barrier and reach the SQM core. They will be converted to SQM and
the surface of the SQM core will be heated, which further reduces the electric field and hence the
gap width \cite{1995PhRvD..51.1440K}. Consequently, a faster tunneling penetration is stimulated
by the decreased gap width. This is a positive feedback and the SS crust will finally collapse
completely on a free-fall timescale of $\sim 0.1\ \rm{ms}$.

Although the detailed mechanism for maintaining a strong magnetic field in various compact stars
is still largely uncertain, it is believed that SSs can also have a strong magnetic field.
When the crust of an SS breaks and falls into the SQM core, the magnetic field lines in the
crust will be dragged into the core due to Alfv\'{e}n's frozen effect. A fraction of the magnetic energy
originally embedded in the crust will be transferred to radiation.
In fact, after the collapse of the SS crust, the magnetic field lines near the
polar cap region will be disturbed and twisted because of differential rotation and/or magnetic instabilities \cite{2013ApJ...776L..39K}.
Hence transient dissipation processes such as magnetic reconnection may be triggered \cite{1995MNRAS.275..255T}, and the magnetic energy will be released on a short timescale.

Theoretically, the limiting interior
magnetic field is of the order of $B_{\rm{max}} \sim 10^{18}\ M_{1} r_{6}^{-2} \ \rm{G}$, where the SS
is assumed to have a mass of $M  = M_{1} \ M_{\odot}$ and a radius of $r = r_{6} \ 10^{6}\ \rm{cm}$ \cite{1991ApJ...383..745L}.
The convention $Q_{x} = Q/10^{x}$ in cgs units is adopted hereafter. Pulsars with surface magnetic fields up to
$\sim 10^{14}\ \rm{G}$ have been reported, and there is no ``smoking-gun'' evidence to identify them as
neutron stars or SSs \cite{1998Natur.393..235K}. It is reasonable to postulate that some SSs could
have a surface magnetic field as strong as $B_{\rm{S}} \sim 10^{14}\ \rm{G}$ in the polar cap region, where the
field should be the strongest for a dipole configuration.
A dipole field approximation $ B(R) \approx B_{\rm {S}} \times (R/r)^{-3} $ is applied in our paper, where $R$ is the distance from the SS center.
The total magnetic energy stored in the crust
can be expressed as $E_{\rm{B}} \sim 4 \pi r^{2} l \times B_{\rm{S}}^{2}/8 \pi \sim 5 \times 10^{43}\ B_{\rm{S,14}}^{2} r_{6}^{2} l_{4}\ \rm{erg}$.

In our framework, we believe that an FRB is produced mainly from the polar cap region,
so not all the $E_{B}$ energy is available for generating the FRB. The angular size of the polar-cap region is
approximately $\theta_{\rm{cap}} \sim 1.45\times 10^{-2} \ P_{0}^{-1/2} r_{6}^{1/2}$, where $P$ is the rotation
period of the SS. The magnetic energy included in the polar region can then be estimated as
$E_{\rm{B,cap}} \sim E_{\rm{B}} \times \pi \theta_{\rm{cap}}^{2}/4 \pi
\sim 2.6 \times 10^{39}\  P_{0}^{-1} B_{\rm{S,14}}^{2} r_{6}^{3} l_{4}\ \rm{erg}$.
We suppose that the radio emission is restricted in an area with the solid angle of $4 \pi f$,
where $f$ is a parameter characterizing the beaming fraction.
The energy needed to produce an FRB is then $\sim f E_{\rm{FRB}}$, where $E_{\rm{FRB}}$ is
the isotropic FRB energy. Assuming $\eta$ to be the fraction of $E_{\rm{B,cap}}$ that is emitted, it can be
calculated as $\eta \sim f E_{\rm{FRB}}/ E_{\rm{B,cap}} \sim 0.4\ f P_{0} B_{\rm{S,14}}^{-2} r_{6}^{-3} l_{4}$.
If we take $f \sim 0.1$ and $P \sim 1\ \rm{s}$ \cite{1990SvAL...16..410K}, $\eta$ can be reckoned as $\sim 0.04$.

In short, as long as a small fraction of the magnetic energy conserved in SS polar cap
regions is transferred into radiation during the collapse
process, the energy is adequate for an FRB.
FRBs should be connected with coherent emission mechanism for their extremely
high brightness temperature \cite{ 2014PhRvD..89j3009K}.
We will discuss the detailed radiation process below.

\section{Emission mechanism} \label{3}

\begin{figure}[htbp]
	\centering\includegraphics[width=3.5in]{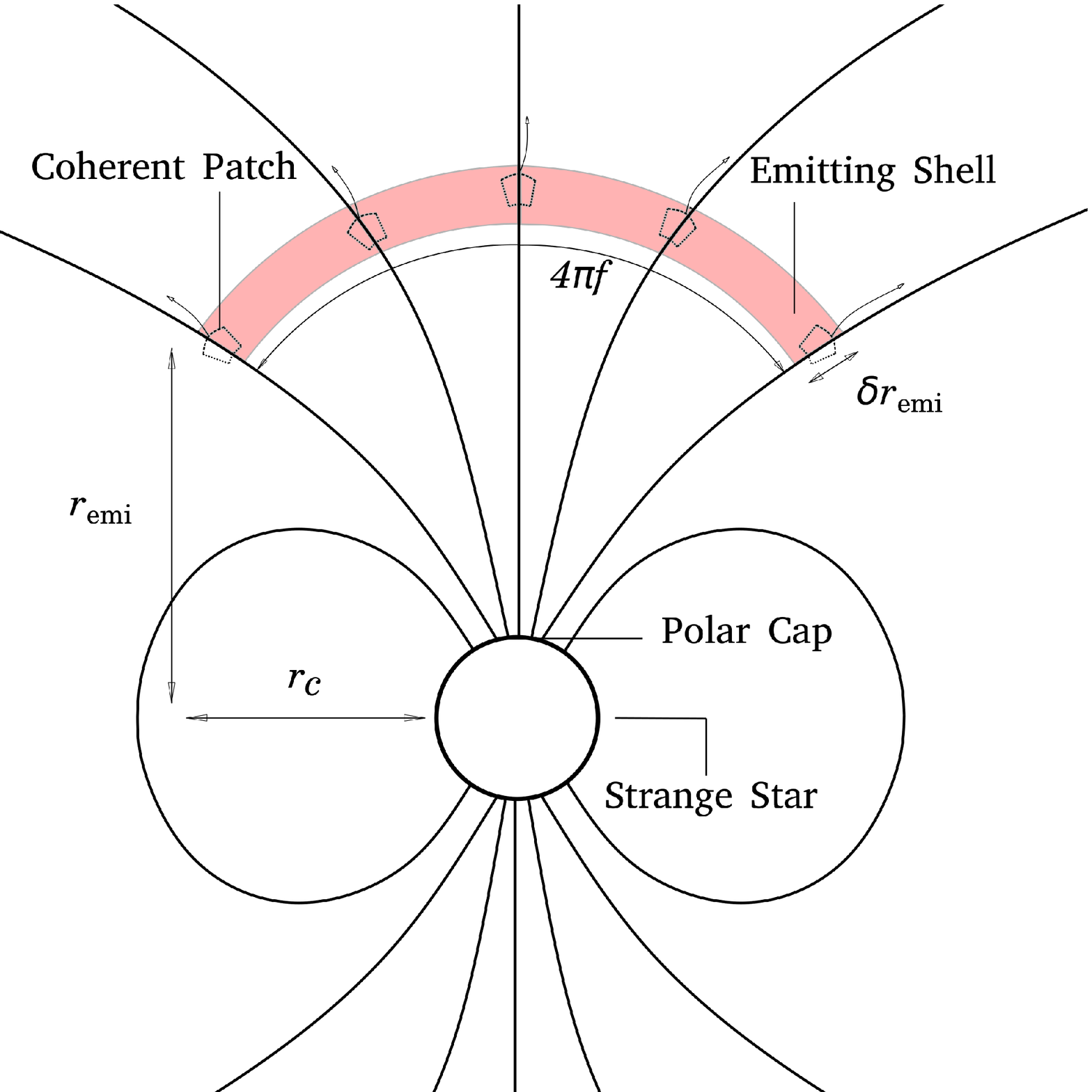}
	\caption{
		A schematic illustration of how an FRB is generated after the collapse of the strange star crust.
		Electrons are accelerated to relativistic velocities and expand along the magnetic field lines to form a shell.
		Coherent emission at radio wavelength is produced when the shell radius reaches $r_{\rm{emi}}$.
		}
	\label{collapse_sketch}
\end{figure}

After the crust collapse, the SS becomes hot and bare.
It turns into a powerful source of electrons and positrons ($e^{+} e^{-}$) pairs.
These $e^{+} e^{-}$ pairs are created in an extremely strong electric field \cite{1998PhRvL..81.4775U}.
Near the polar cap region where the magnetic field energy is released,  electrons and positrons will be accelerated to ultra-relativistic speeds \cite{1975ApJ...196...51R, 1977MNRAS.179..189B}.
These particles coast along the magnetic field lines and form a shell with a thickness of $\delta r_{\rm{emi}} \approx c \delta t$.
This process is illustrated in Figure \ref{collapse_sketch}.

Curvature radiation will be produced when electrons in the shell are streaming along parallel magnetic field lines.
For simplicity, we postulate that all electrons are moving with almost exactly the same
velocity (i.e., with the corresponding Lorentz factor of $\gamma$), and $\delta r_{\rm{emi}}$ remains roughly constant in the observer frame as long as $\delta r_{\rm{emi}} >  \gamma^{-2} r_{\rm{emi}}$ like the fireballs in the cases of gamma-ray bursts \cite{ 1992MNRAS.258P..41R}.
Let $r_{\rm{emi}}$ be the emission radius of the shell and $r_{\rm{c}}$ be the curvature
radius of the magnetic field lines, then the shell volume is $V_{\rm{emi}} \approx 4 \pi f r_{\rm{emi}}^{2} \delta r_{\rm{emi}}$.
The patch in which electrons radiate coherently has a characteristic radial size $\lambda \approx c/\nu_{\rm{c}}$
and a solid angle $4/\gamma^{2}$ according to beaming effects.
Such a patch has a volume of $V_{\rm{coh}} \approx (c/ \nu_{\rm{c}}) \times (4/\gamma^{2}) r_{\rm{emi}}^{2}$ \cite{ 2017MNRAS.468.2726K, 2013ApJ...776L..39K}.
In the emission region, there are $N_{\rm{pat}} \approx V_{\rm{emi}}/V_{\rm{coh}}$ coherent patches totally,
and each patch has $N_{\rm{coh}} \approx n_{\rm{e}} \times V_{\rm{coh}}$ electrons in it, where $n_{\rm{e}}$ is the electron number density.
The frequency of curvature emission is $\nu_{\rm{c}} \approx  \gamma^{3}(3 c/4 \pi r_{c}) \approx \nu_{\rm{FRB}}$.
The total coherent curvature emission luminosity from these electrons can be expressed as $L_{\rm{total}} \approx (P_{\rm{e}} N_{\rm {coh}}^{2}) \times N_{\rm {pat}}$,
where  $P_{\rm {e}}  = 2 \gamma^{4} e^{2} c / 3 r_{\rm {c}}^{2}$ is the emission power of a single electron \cite{2013ApJ...776L..39K}.

{
According to \cite{1977MNRAS.179..189B}, the coherent radiation peaks at the places where the relativistic
plasma energy density just exceeds the dipolar field energy density.
With the plasma pressure of $\rho_{\rm {P}}(R) \approx n_{\rm{e}} \gamma m_{\rm {e}} c^{2}$,
the magnetic energy density of $\rho_{\rm{M}}(R) \approx B^{2}(R) / 8 \pi$, and assuming $r_{\rm{emi}} \sim r_{\rm{c}}$,
the emission radius can be estimated as
\begin{equation}
r_{\rm{emi}} \sim 0.6 \times 10^{10}\ \left(f_{-1}^{3} L_{\rm{total,43}}^{-3}   \delta t_{-3}^{3} \nu_{\rm{FRB,9}}^{-1} B_{\rm{S,14}}^{12} r_{6}^{36}\right)^{1/25}\ \rm{cm}.
\end{equation}
}
On other hand, the electron Lorentz factor $\gamma$ can be derived as
\begin{equation}\label{gamma}
%\gamma \sim 520 \ \nu_{\rm {FRB},9}^{1/3} r^{1/3}_{\rm{c},9}.
\gamma \sim 1120 \ \nu_{\rm {FRB},9}^{1/3} r^{1/3}_{\rm{c},10}.
\end{equation}
The total coherent curvature emission luminosity from these electrons can be expressed as  \cite{2013ApJ...776L..39K}
\begin{eqnarray}
L_{\rm{total}} &\approx& (P_{\rm{e}} N_{\rm {coh}}^{2}) \times N_{\rm {pat}} \nonumber\\
&\sim& 2.6 \times 10^{42}\ \rm{erg \ s^{-1}}  \nonumber\\
& \times& f n_{\rm {e,7}}^{2} \delta t_{-3} \nu_{\rm {FRB,9}}^{-1/3} r_{\rm {emi,10}}^{4} r_{\rm {c,10}}^{-4/3}.
\end{eqnarray}
Then the observed isotropic FRB luminosity
$L_{\rm{FRB}} \approx f^{-1} L_{\rm{total}} \sim 10^{42 - 43}\ \rm{erg\ s^{-1}}$ is consistent with observations.

{
The typical emission frequency should be larger than the plasma characteristic frequency so that
the radio waves can propagate without being absorbed
\cite{1977MNRAS.179..189B, 2013ApJ...776L..39K}, i.e.,
\begin{equation}
\nu_{\rm{FRB}} >= f_{\rm{pe}} \approx \left(\frac{\gamma n_{\rm{e}} e^{2}}{\pi m_{\rm{e}}}\right)^{1/2}.   \label{f_pe}
\end{equation}
It requires that
\begin{equation}
n_{\rm{e}} <= 1.1 \times 10^{7}\ \nu_{\rm{FRB},9}^{5/3} r_{\rm{c},10}^{-1/3}\ \rm{cm}^{-3}.
\end{equation}
Another requirement is that the induced Compton scattering should not be too strong to
hinder the propagation of the radio wave in the plasma \cite{ 1971Ap&SS..13...56M}.
The Lorentz factor of the relativistic plasma is therefore limited by (see Equation (53) of \cite{2008ApJ...682.1443L})
\begin{eqnarray}
\gamma & >= & 730\ \gamma_{\rm{T}}^{-1/4} \zeta^{-1/8} \nu_{\rm{FRB,9}}^{-1/8}\nonumber \\
& \times&  \left(\frac{F}{1\ \rm{Jy}}\right)^{1/4} \left(\frac{\delta t}{5\ \rm{ms}}\right)^{-3/8} \left(\frac{D}{100\ \rm{Mpc}}\right)^{1/2}.
\end{eqnarray}
Here $\gamma_{\rm{T}} \sim 1$ is the thermal Lorentz factor of the electrons/positrons in the plasma's co-moving frame,
$\zeta$ is the fraction of the plasma energy radiated in radio,
$F$ is the radio flux density, and $D$ is the distance to the source.
Comparing Equation~(6) with Equation~(2), we find that this requirement can be reasonably satisfied.
}

The coherent energy loss rate per electron is $P_{\rm{e}} N_{\rm{coh}} \sim 7.7 \times 10^{6}\ n_{\rm{e,7}} \gamma_{3}^{-1} r_{\rm{emi,10}}^{2} r_{\rm{c,10}}^{-1}\ \rm{erg\ s^{-1}}$.
Meanwhile, electrons will be accelerated by the rotating SS at the rate 
of $2 \pi e B(r_{\rm{emi}}) r_{\rm{emi}} P_{0}^{-1} \sim 3.0 \times 10^{3}\ r_{6}^{3} r_{\rm{emi,10}}^{-2} B_{\rm{S,14}} P_{0}^{-1}\ \rm{erg\ s^{-1}}$ \cite{2013ApJ...776L..39K}, where $B(r_{\rm{emi}})$ is the magnetic field strength at $r_{\rm{emi}}$.
The maximum Lorentz factor during coherent emission thus can be obtained, $\gamma_{\rm{max}} \sim 2.6 \times 10^{6}\ n_{\rm {e,7}}r_{\rm{emi,10}}^{4} r_{\rm{c,10}}^{-1}  r_{6}^{3} B_{\rm{S,14}}^{-1} P_{0}$, which is large enough for FRBs.

%The relation between coherent emission spatial regions and frequency spectrum has been studied by \cite{1977MNRAS.179..189B}.
%The peak in the spectrum corresponds to the places where the relativistic plasma energy density just exceeds the dipolar field energy density.
%The magnetic energy density can be expressed as $\rho_{\rm{M}}(R) \approx B^{2}(R) / 8 \pi$, while the plasma pressure can be evaluated as $\rho_{\rm {P}}(R) \approx n_{\rm{e}} \gamma m_{\rm {e}} c^{2}$.
%Thus we have the pressure balanced point radius
%\begin{equation}
%	 	r_{\rm{b}}  \sim 0.6\  \times 10^{10}\ \emph{n}_{\rm{e},7}^{-1/4} \nu_{\rm{FRB},9}^{1/12} \emph{r}_{\rm{c,10}}^{-1/12} \emph{B}_{\rm{S},14}^{1/3} \emph{r}_{\rm{SS},6}\ \rm{cm}
%\end{equation}
%and $r_{\rm{emi}}$ shall be slightly larger than $r_{\rm{b}}$.

\section{Counterparts in Other Wavebands} \label{4}
No counterparts of FRBs have been discovered in other wavebands yet{ , except for the repeating FRB, FRB 121102 \cite{2018Natur.553..182M}}.
The collapse of a SS crust might result in electromagnetic radiation besides radio \cite{2015MNRAS.447..246P}.
It's intriguing to check whether the emissions in other wavebands is strong or not in our model.

There are mainly two types of emission from a bare SS surface, thermal radiation and $e^{+} e^{-}$
pair emission \cite{2001ApJ...550L.179U}. The plasma frequency for an SQM object can be written as
\begin{equation}
	\omega_{\rm {p}} = \left(\frac{8 \pi}{3} \ \frac{e^{2}c^{2}n_{\rm {b}}}{\mu} \right)^{1/2},
\end{equation}
where $n_{\rm {b}}$ is the baryon number density of SQM and $\mu \simeq \hbar c (\pi^{2} n_{\rm {b}})^{1/3}$ \cite{ 1986ApJ...310..261A}.
According to the plasma dispersion relationship, propagating modes for electromagnetic waves with $\omega < \omega_{\rm{P}}$ do not exist.
For typical SQM $n_{\rm {b}} \simeq (1.5 \sim 2)n_{0}$, where $n_{0}  \simeq 1.7 \times 10^{38}\ \rm{cm^{-3}}$ is normal nuclear matter density, we expect $\hbar \omega_{p} \simeq 20-25 \ \rm {MeV}$.
Thus SSs with a surface temperature $T_{\rm{S}}$ below $2\ \times\ 10^{10}\ \rm{K}$ are very poor radiators for blackbody radiation.
However, the thermal radiation luminosity increases sharply and becomes the chief emission form if $T_{\rm{S}} \geq 5\ \times\ 10^{10}\ \rm{K}$ \cite{1991NuPhS..24..139H,1986ApJ...310..261A}.
The energy flux per unit surface in thermal photons is
$F_{\rm{eq}} = \frac{\hbar}{c^{2}} \int_{\omega_{\rm{p}}}^{\infty}
\frac{\omega (\omega^{2} -\omega_{\rm{p}}^{2}) g(\omega)}{\exp(\hbar \omega / k_{\rm{B}}) - 1} d \omega $,
where $g(\omega) = \frac{1}{2 \pi^{2}} \int_{0}^{\pi/2} D(\omega,\theta) \sin\theta \cos\theta d\theta$,
$\kappa_{\rm{B}}$ is the Boltzmann constant, $D(\omega, \theta)$ is the coefficient of radiation transmission
from SQM to the vacuum, $D = 1 - (R_{\perp} + R_{\parallel})$, and $R_{\perp}
= \sin^{2} (\theta - \theta_{0})/\sin^{2} (\theta + \theta_{0})$, $R_{\parallel}
= \tan^{2} (\theta - \theta_{0})/\tan^{2} (\theta + \theta_{0})$,
$\theta_{0} = \arcsin[\sin \sqrt{1-(\omega_{p} /\omega)^{2}}]$ \cite{ 2001ApJ...550L.179U}.

As pointed out by \cite{1998PhRvL..80..230U}, the Coulomb barrier outside the SS surface is a very powerful
source of $e^{+}e^{-}$ pairs, where the electronic field  $\sim 5\ \times 10^{17} \ \rm{V \ cm^{-1}}$ is
tens of times higher than the critical vacuum polarization field $E_{\rm {cr}} = m^{2} c^{3} /e \hbar \simeq 1.3 \ \times 10^{17}\ \rm {V \ cm^{-1}}$.
$e^{+}e^{-}$ pairs are created with the mean particle energy of $\varepsilon_{\pm} \simeq m_{\rm {e}} c^{2} + k T_{\rm {S}}$.
The flux of pairs is
\begin{equation}
	f_{\pm} \simeq 10^{39.2} \left(\frac{T_{\rm{S}}}{10^{9}\ \rm{K}} \right) \ \exp\left(-\frac{11.9 \times 10^{9}\ \rm{K}}{T_{\rm{S}}}\right) J\left(\xi\right)\ \rm{s^{-1}},
\end{equation}
where
\begin{equation}
	J\left(\xi\right) = \frac{1}{3} \frac{\xi^{3} \ln(1+2\xi^{-1})}{(1+0.074 \xi)^{3}} + \frac{\pi^{5}}{6} \frac{\xi^4}{(13.9 + \xi)^{4}},
\end{equation}
and $\xi \simeq (2 \times 10^{10}\ \rm{K})/T_{\rm {S}}$ \cite{ 2001ApJ...550L.179U}.
The energy flux per unit surface in $e^{+} e^{-}$ pairs is $F_{\pm} \simeq \varepsilon_{\pm} f_{\pm}$.

Since the radiation features of a bare SS are determined by $T_{\rm{S}}$, it is crucial to study the distribution of temperature and how it evolves.
The heat transfer equation for SSs under the plane-parallel approximation is \cite{ 2001PhRvL..87b1101U,  1982AnPhy.141....1I,  1993PhRvD..48.2916H}
\begin{equation}
	C_{q} \frac{\partial T}{\partial t} = \frac{\partial}{\partial x}
	\left( K_{\rm {c}} \frac{\partial T}{\partial x}\right) - \varepsilon_{\nu},
\end{equation}
where $C_{\rm {q}} \simeq 2.5 \times 10^{20}\ (n_{\rm{b}}/n_{0})^{2/3}
T_{9}\ \rm{erg\ cm^{-3}\ K^{-1}} $ is the specific heat for SQM per unit volume,
$K_{\rm{c}} \simeq 6 \times 10^{29}\ \alpha_{\rm{c}}^{-1} (n_{\rm{b}}/n_{0})^{2/3}\ \rm{erg\ cm^{-1}\ K^{-1}}$
is the thermal conductivity,
$\epsilon_{\rm{\nu}} \simeq 2.2 \times 10^{26}\ \alpha_{\rm{c}} Y_{\rm{e}}^{1/3} (n_{\rm{b}}/n_{0}) T_{9}^{6}\
\rm{erg\ cm^{-3}\ s^{-1}}$ is the neutrino emissivity, $n_{\rm{b}} \sim 2 n_{0}$ is the SQM baryon number density,
$\alpha_{\rm{c}} = g^{2}/4 \pi \sim 0.1$ is the QCD fine structure constant with $g$ being
the quark-gluon coupling constant, and $Y_{e} =n_{\rm{e}}/n_{\rm{b}} \sim 10^{-4}$ is the number ratio between electrons and baryons.
Due to thermal conductivity, the heat flux is
\begin{equation}
	q = -K_{\rm{c}} dT/dx.
\end{equation}
The boundary condition is $q \simeq - F_{\pm} - F_{\rm{eq}}$.

We now proceed to calculate the evolution of the radiation luminosity after the collapse of an SS crust.
Our calculations are done for an SS with an initial surface and crust temperature of $T_{\rm {S} 0} \sim 3 \times 10^{7} \ \rm{K}$ \cite{  1991PhRvL..66.2425P, 1997ApJ...481L.107U}, and a crust mass of $M_{\rm{c}} \sim 3 \times 10^{-6}\ M_{\odot}$.
There are mainly two kinds of energy released during the collapse:
(1) gravitational energy of the crust ($\sim 0.002\ M_{c} c^{2}$) due to its radical movement to the SQM surface, and
(2) deconfinement energy ($\sim 0.01 - 0.03\ M_{\rm{c}} c^{2}$) due to the conversion of the crust material to SQM \cite{ 1996PhRvL..77.1210C}.
As a sum, we can take the typical total energy from these two reserviors  as roughly
$Q \sim 0.02\ M_{\rm{c}} c^{2} \sim 1 \times 10^{47}\ \rm{erg}$ for $M_{\rm{c}} \sim 3 \times 10^{-6}\ M_{\odot}$.
After the collapse, the crust is transferred to SQM with a density
of $\sim 5 \times 10^{14}\ \rm{g\ cm^{-3}}$ \cite{ 1986ApJ...310..261A} and the interaction is restricted in a thickness of $l^{*} \sim 1\ \rm{cm}$.
Assuming the actual combustion mode is detonation \cite{ 1987PhLB..192...71O,  1988PhLB..213..516H, 1991NuPhS..24..144H},
the timescale of conversion is rather small and the interior SS temperature can be treated as uninfluenced while
the surface layer is hot and isothermal.
With $C_{\rm {q}} \sim 4 \times 10^{11}\ T\ \rm{erg\ cm^{-3}\ K^{-1}} $,
the temperature of the heated layer can be estimated as $T_{\rm{S}}^{*}
\sim 2 \times 10^{11}\ \rm{K}$.
The initial temperature distribution can be expressed as
\begin{equation}
	T(t = 0,x) \simeq \left\{
	\begin{aligned}
	&T_{\rm{S}}^{*},\ 0 < x < l^{*}&\\
	&T_{\rm{S} 0},\ x \geq l^{*},&
	\end{aligned}
	\right.
\end{equation}
where $t$ is the time after the collapse and $x$ is a space coordinate measuring the depth below the SS surface.

\begin{figure}[htbp]
	\centering\includegraphics[width=3.5in]{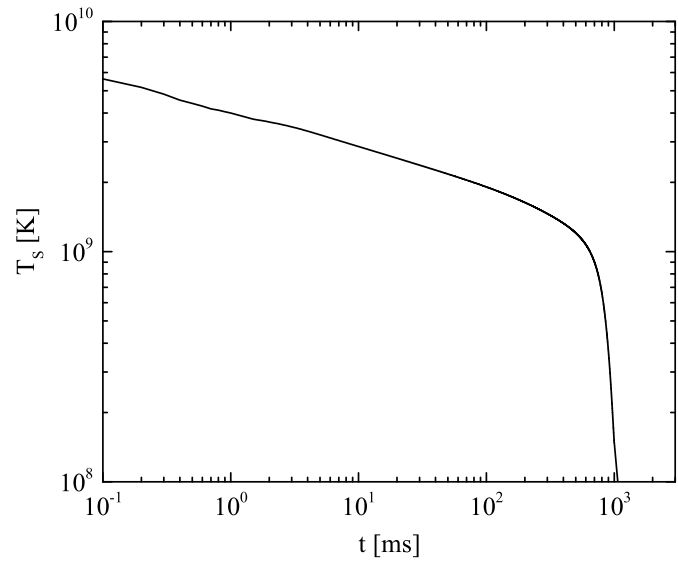}
	\caption{Evolution of the surface temperature of a bare strange star, $T_{\rm{S}}$, after the crust collapses.}
	\label{temperature}
\end{figure}

\begin{figure}[htbp]
	\centering\includegraphics[width=3.5in]{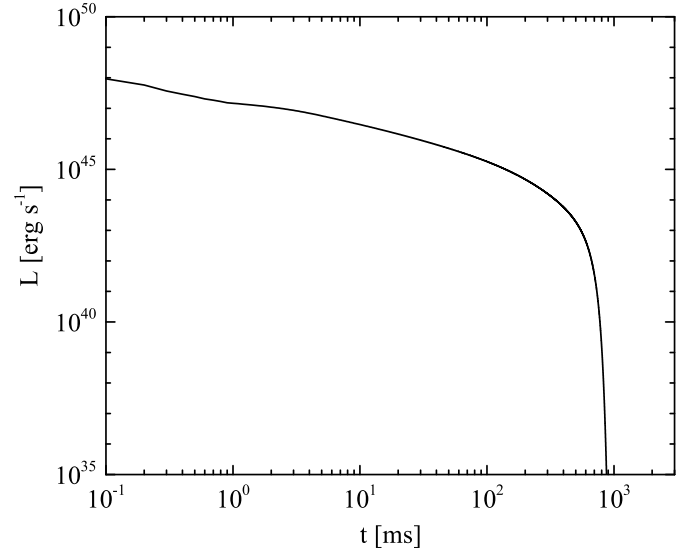}
	\caption{Total luminosity, $L$, as a function of time after the crust collapses. In this figure,
             $L = L_{\pm} + L_{\rm{eq}}$, where $L_{\pm}$ and $ L_{\rm{eq}}$ are the luminosities in
             thermal photons and $e^{+} e^{-}$ pairs, respectively.}
	\label{luminosity}
\end{figure}

Combing these postulations and approximations, we have performed numerical calculations to solve the cooling process.
Figure \ref{temperature} shows the evolution of the surface temperature as a function of time.
The total luminosity including both photons and electrons/positrons,
$L = L_{\rm{eq}} + L_{\pm} = 4 \pi r^{2} (F_{\rm{eq}} + F_{\pm})$,  has also been calculated and the result is shown in Figure \ref{luminosity}.
We should note that when $L_{\pm}$ is very high, most of the pairs will annihilate into photons near the SS surface due to the high pair density.
Hence the emerging emission consists mostly of photons.
The photon spectrum is roughly a blackbody with a high energy ($> 100\ \rm{keV}$) tail \cite{2003MNRAS.343L..69A} 
since the outgoing pairs and photons are very likely in thermal equilibrium \cite{2002ApJ...565..163I}.
Note that bare SSs are bounded by strong interactions rather than the gravity, so the luminosity can safely exceed the Eddington limit \cite{1986ApJ...310..261A}.

The luminosity distance is $d_{\rm{L}} \sim $ a few hundred $ \rm{Mpc}$ at redshift of $z \sim 0.5 - 1$.
The hot bare SS radiates at an extremely high luminosity just after the collapse and it cools down rapidly.
The surface layer will become cold soon, making the radiation power decrease quickly.
Since $T_{\rm{S}} \sim $ a few $10^{9}\ \rm{K}$ during the emission, the typical photon energy can be estimated
as $\sim 100\ \rm{keV}$ for a blackbody spectrum. According to our calculations, the emitted energies are
roughly $9.6 \times 10^{44}\ \rm{erg}$, $1.6 \times 10^{45}\ \rm{erg}$, and $1.7 \times 10^{45}\ \rm{erg}$
in the first $10\ \rm{ms}$, $100\ \rm{ms}$, and $1000\ \rm{ms}$ after the collapse, respectively.
If averaged over $100\ \rm{ms}$, the typical luminosity is $L \sim 2 \times 10^{46}\ \rm{erg\ s^{-1}}$.
Noting that the expected radiation flux is then $10^{-8}\ L_{46}\ (d_{\rm{L}} /100\ \rm{Mpc})^{-2}\ \rm{erg\ cm^{-2}\ s^{-1}}$,
this luminosity will be too low to trigger Swift BAT, whose threshold is $\sim 10^{-7}\ \rm{erg\ cm^{2}\ s^{-1} }$
in $15 - 150\ \rm{keV}$ \cite{ 2009A&A...500.1193L}.
Also, the derived peak photon flux of $\sim 0.1\ (d_{\rm{L}} /100\ \rm{Mpc})^{-2}\ \rm{photons\ cm^{-2}\ s^{-1}}$
is lower than the trigger threshold of Fermi GBM, $ 0.74\ \rm{photons\ cm^{-2}\ s^{-1}}$ \cite{ 2009ApJ...702..791M}.
Additionally, the extremely small radiation timescale of about $0.1\ \rm{s}$ makes the observation even more difficult.
Although convection process may reheat the SS surface and increase its emission slightly \cite{ 1998PhRvL..81.4775U},
the light curve within initial $0.1\ \rm{s}$ will not be influenced significantly and no afterglow could
be detected in X-rays or gamma-rays by current high energy detectors.

\section{Summary and Discussion} \label{5}
In summary, we propose that FRBs may be generated from the collapses of strange star crusts.
During the collapse, a fraction of magnetic energy is transferred to accelerate electrons and
positions in the polar cap region to relativistic velocities. The accelerated electrons expand
along magnetic field lines to form a shell. At the radius of $r_{\rm{emi}} \sim 10^{10}\ \rm{cm}$,
coherent emission in radio bands will be produced, giving birth to the observed FRB.
At the same time, the emission in X-ray and $\gamma$-ray bands is too faint to be detected by current detectors.

It is argued that the magnetic field of an SS is influenced by spacial temperature variations \cite{2001A&A...371..963X}.
The SQM surface will be heated up to a high temperature of $\sim 10^{11}\ \rm{K}$ and cool down drastically via
the production of electron-positron pairs and neutrinos \cite{1998PhRvL..80..230U, 2001PhRvL..87b1101U}.
A thin layer surface will be even colder than inside during the cooling process.
A cold dense layer forms and then the temperature distribution seems unstable with respect to 
convective disturbances \cite{1998PhRvL..81.4775U}.
The small-scale buoyant convection induced by temperature gradient may amplify the magnetic field 
due to the interaction with differential rotation through fast dynamos processes \cite{2001A&A...371..963X}.
The amplified magnetic field lines may emerge from the stellar interior, where the fields can 
be several order-of-magnitude larger \cite{2011LRSP....8....6S}.
A fraction of heat may be transfered to magnetic field energy and then also contribute to the emission.
In other words, the SS surface magnetic field strength may increase significantly after the collapse, 
thus SSs originally having a weaker magnetic field may also produce FRBs.
The effect of convection needs further investigation in the future.

It is an interesting question whether FRBs generated from SS crust collapses can repeat or not.
The crux is to determine whether a bare SS can re-construct its crust through accretion or other ways.
In fact, most SS formation models are explosive \cite{ 1991ApJ...375..209H, 2001APh....15..101X, 2013PhRvD..87j3007P},
thus a newly-born SS is bare and how a normal matter crust forms still needs to be investigated.
The free-fall kinetic energy for a proton onto SS surface is $E_{\rm{p}} \simeq \frac{G M m_{\rm {p}}}
{R \sqrt{1 - 2\ G M / R c^{2}}} \nonumber \approx 138\ M_{1} R_{6}^{-1} (1 + 0.2\ M_{1} R_{6}^{-1}) \
\rm {MeV}$ \cite{1987PhLB..192...71O}, where $m_{{\rm{p}}}$ is the proton mass.
If the SS is non-rotational and not magnetized, and the accretion is isotropic,
then the free-fall energy will be high enough to overcome the electrostatic potential barrier
of $e V \simeq 20 \ \rm{MeV}$ outside the SQM surface \cite{1986ApJ...310..261A}.
So, SS crusts cannot be built in such scenarios.
However, the falling material will have a non-zero angular momentum when approaching 
the surface of rotational magnetized SSs due to the magnetic freezing.
These matter will finally hit the SS surface obliquely at a typical incidence angle with 
cotangent $\sim 0.05$ \cite{1987PhLB..192...71O, 1985ApJ...297..548K}.
Hence the accreted matter has a longer interaction with the electric field and
the radial velocity will be reduced by friction and radiation.
Considering this effect, an envelope built from the accreted material will cover the whole SQM core gradually.
If the accretion continues, the crust will finally reach the critical mass via accretion and then collapse.

{
In this case, the number density of particles near the SS may be rather high due to the existence of the accretion flow.
Thus the interaction between the expanding shell and the accretion flow needs to be considered. 
Assuming that the accretion is isotropic, the number density of the accretion flow at the emission radius is $n_{\rm{a}} \approx 2 \dot{M} / \left[4 \pi r_{\rm{emi}}^2 m_{\rm{p}} (2 G M / r_{\rm{emi}})^{1/2}\right]
 \sim 1.6 \times 10^{5}\ (\dot{M} /10^{-15}\ M_{\odot}\ \rm{yr}^{-1})$ $ (M/1.4\ M_{\odot}) r_{\rm{emi}, 10}^{-3/2}\ 
 \rm{cm}^{-3} $, where $\dot{M}$ is the accretion rate.
If the accreting rate is high, the outgoing shell may be disrupted by the falling gas and then no FRB could be generated. 
However, as long as $\dot{M}$ is less than a critical value of $\dot{M}_{\rm{cr}} \approx 10^{-15}\ M_{\odot}\ \rm{yr}^{-1}$,
we notice that $n_{\rm{a}}$ will be much smaller than the number density of electrons in the expanding shell ($n_{\rm{e}}$). 
Then the influence of the accretion flow can be negligible.}

{\bf
For SSs formed in explosive events \cite{1991ApJ...375..209H, 2001APh....15..101X, 2013PhRvD..87j3007P},
the compact stars may receive a ``kick'' if these drastic events are asymmetric \cite{2012MNRAS.423.1805N, 2016MNRAS.461.3747B}.
The typical kick velocity, $v$, ranges from 200 to 400 $\rm{km\ s^{-1}}$, with the highest value even 
in excess of $1000\ \rm{km\ s^{-1}}$ \cite{1994Natur.369..127L, 2006ApJ...643..332F}.
SSs with a kick velocity would accrete ambient matter at a rate of $\dot{M} \approx 2 \pi \alpha (G M)^2 v^{-3} \rho \sim 2 \times 10^{-15}\ M_{1}^{2} \rho_{-21} v_{7}^{-3}\ M_{\odot}\ \rm{yr}^{-1}$, where $\alpha \sim 1.25$ is a numerical constant and 
$\rho$ is the ambient matter density \cite{1952MNRAS.112..195B}.
In this case, the typical $\dot{M}$ would not exceed $\dot{M}_{\rm{cr}}$ significantly.
Therefore, the expanding electron/positron shell can expand without being destroyed by the accretion flow in our scenario.
For this kind of accretion, the reconstruction timescale of the crust can be roughly estimated 
as $\tau_{\rm{rec}} \approx M_{\rm{c}}/\dot{M}_{\rm{cr}} \sim 10^{9}\ \rm{yr}$.}

Owing to this long reconstruction timescale, multiple FRB events from the same source seem not likely to happen
in our scenario. Our model thus is more suitable for explaining the non-repeating FRBs, and the repeating $\rm{FRB} \ 121102 $
may be produced via other mechanisms \cite{ 2018ApJ...854L..12P}.
However, we should also note that during the collapse process, if only a small portion (in the polar cap
region) of the crust falls onto the SQM core while the other portion of the crust remains stable, then the
rebuilt timescale for the crust can be markedly reduced and repeating FRBs would still be possible.
Further detailed studies on the crust collapse thus still need to be conducted.

The event rate of FRBs is as high as $2 \times 10^{3}\ \rm{sky^{-1}\ day^{-1}}$ \cite{2018MNRAS.475.1427B, 2017RAA....17....6L}.
There are roughly $\sim 10^{9}$ galaxies within the redshift range of $z \leq 1$ and the total number of pulsars per galaxy
is $\sim 10^{8}$ on average \cite{ 1996ApJ...457..834T}. The event rate for SS crust collapse can be
estimated as $\sim 2 \times 10^{3}\ (\tau_{\rm{rec}}/10^{9}\  \rm{yr})^{-1}\ $$f f_{\rm{S},-3}$,
where $f$ is the beaming factor of radio emission, and $f_{\rm{S}}$ is the number ratio of SSs with compatible conditions to generate FRBs over pulsars.
Still, we would like to remind that FRBs may be of multiple origin and our model may only contribute a portion of them.
Further observations and larger samples in the future would help to solve the enigma finally. 

We appreciate valuable comments and constructive suggestions from the anonymous referee.
This work is supported by
the National Natural Science Foundation of China (Grant No. 11473012),  %HYF
the National Basic Research Program (\textquotedblleft
973\textquotedblright\ Program) of China (Grant No. 2014CB845800), %HYF
the National Postdoctoral Program for Innovative Talents (Grant No. BX201700115),
the China Postdoctoral Science Foundation funded project (Grant No. 2017M620199),
and the Strategic Priority Research Program \textquotedblleft Multi-waveband gravitational wave
Universe\textquotedblright\ (Grant No. XDB23040000) of the Chinese Academy of Sciences. % HYF

\begin{thebibliography}

\bibitem[Abe et al.(1984)]{1984PhRvD..30....1A} Abe, K., Bacon, T.~C., Ballam, J., et al.\ 1984, \prd, 30, 1


\bibitem[Aksenov et al.(2003)]{2003MNRAS.343L..69A} Aksenov, A.~G., Milgrom, M., \& Usov, V.~V.\ 2003, \mnras, 343, L69


\bibitem[Alcock et al.(1986)]{1986ApJ...310..261A} Alcock, C., Farhi, E., \& Olinto, A.\ 1986, \apj, 310, 261


\bibitem[Barrau et al.(2014)]{2014PhRvD..90l7503B} Barrau, A., Rovelli, C., \& Vidotto, F.\ 2014, \prd, 90, 127503


\bibitem[Benford \& Buschauer(1977)]{1977MNRAS.179..189B} Benford, G., \& Buschauer, R.\ 1977, \mnras, 179, 189


\bibitem[Bhandari et al.(2018)]{2018MNRAS.475.1427B} Bhandari, S., Keane, E.~F., Barr, E.~D., et al.\ 2018, \mnras, 475, 1427


\bibitem[Bondi(1952)]{1952MNRAS.112..195B} Bondi, H.\ 1952, \mnras, 112, 195


\bibitem[Bray \& Eldridge(2016)]{2016MNRAS.461.3747B} Bray, J.~C., \& Eldridge, J.~J.\ 2016, \mnras, 461, 3747


\bibitem[Cheng \& Dai(1996)]{1996PhRvL..77.1210C} Cheng, K.~S., \& Dai, Z.~G.\ 1996, Physical Review Letters, 77, 1210


\bibitem[Chmaj et al.(1991)]{1991NuPhS..24...40C} Chmaj, T., Haensel, P., \& S{\l}omi{\'n}ski, W.\ 1991, Nuclear Physics B Proceedings Supplements, 24, 40


\bibitem[Cordes \& Lazio(2002)]{2002astro.ph..7156C} Cordes, J.~M., \& Lazio, T.~J.~W.\ 2002, arXiv:astro-ph/0207156


\bibitem[Dai et al.(2016)]{2016ApJ...829...27D} Dai, Z.~G., Wang, J.~S., Wu, X.~F., \& Huang, Y.~F.\ 2016, \apj, 829, 27


\bibitem[Falcke \& Rezzolla(2014)]{2014A&A...562A.137F} Falcke, H., \& Rezzolla, L.\ 2014, \aap, 562, A137


\bibitem[Farhi \& Jaffe(1984)]{1984PhRvD..30.2379F} Farhi, E., \& Jaffe, R.~L.\ 1984, \prd, 30, 2379


\bibitem[Faucher-Gigu{\`e}re \& Kaspi(2006)]{2006ApJ...643..332F} Faucher-Gigu{\`e}re, C.-A., \& Kaspi, V.~M.\ 2006, \apj, 643, 332


\bibitem[Geng \& Huang(2015)]{2015ApJ...809...24G} Geng, J.~J., \& Huang, Y.~F.\ 2015, \apj, 809, 24


\bibitem[Geng et al.(2015)]{2015ApJ...804...21G} Geng, J.~J., Huang, Y.~F., \& Lu, T.\ 2015, \apj, 804, 21


\bibitem[Gu et al.(2016)]{2016ApJ...823L..28G} Gu, W.-M., Dong, Y.-Z., Liu, T., Ma, R., \& Wang, J.\ 2016, \apjl, 823, L28


\bibitem[Haensel et al.(1991)]{1991ApJ...375..209H} Haensel, P., Paczynski, B., \& Amsterdamski, P.\ 1991, \apj, 375, 209


\bibitem[Haensel et al.(1986)]{1986A&A...160..121H} Haensel, P., Zdunik, J.~L., \& Schaefer, R.\ 1986, \aap, 160, 121


\bibitem[Haensel(1991)]{1991NuPhS..24...23H} Haensel, P.\ 1991, Nuclear Physics B Proceedings Supplements, 24, 23


\bibitem[Haensel \& Zdunik(1991)]{1991NuPhS..24..139H} Haensel, P., \& Zdunik, J.~L.\ 1991, Nuclear Physics B Proceedings Supplements, 24, 139


\bibitem[Heiselberg et al.(1991)]{1991NuPhS..24..144H} Heiselberg, H., Baym, G., \& Pethick, C.~J.\ 1991, Nuclear Physics B Proceedings Supplements, 24, 144


\bibitem[Heiselberg \& Pethick(1993)]{1993PhRvD..48.2916H} Heiselberg, H., \& Pethick, C.~J.\ 1993, \prd, 48, 2916


\bibitem[Horvath \& Benvenuto(1988)]{1988PhLB..213..516H} Horvath, J.~E., \& Benvenuto, O.~G.\ 1988, Physics Letters B, 213, 516


\bibitem[Huang \& Lu(1997)]{1997A&A...325..189H} Huang, Y.~F., \& Lu, T.\ 1997, \aap, 325, 189


\bibitem[Iwamoto(1982)]{1982AnPhy.141....1I} Iwamoto, N.\ 1982, Annals of Physics, 141, 1


\bibitem[Iwamoto \& Takahara(2002)]{2002ApJ...565..163I} Iwamoto, S., \& Takahara, F.\ 2002, \apj, 565, 163


\bibitem[Kashiyama et al.(2013)]{2013ApJ...776L..39K} Kashiyama, K., Ioka, K., \& M{\'e}sz{\'a}ros, P.\ 2013, \apjl, 776, L39


\bibitem[Katz(2014)]{2014PhRvD..89j3009K} Katz, J.~I.\ 2014, \prd, 89, 103009


\bibitem[Keane et al.(2012)]{2012MNRAS.425L..71K} Keane, E.~F., Stappers, B.~W., Kramer, M., \& Lyne, A.~G.\ 2012, \mnras, 425, L71


\bibitem[Kettner et al.(1995)]{1995PhRvD..51.1440K} Kettner, C., Weber, F., Weigel, M.~K., \& Glendenning, N.~K.\ 1995, \prd, 51, 1440


\bibitem[Kluzniak \& Wagoner(1985)]{1985ApJ...297..548K} Kluzniak, W., \& Wagoner, R.~V.\ 1985, \apj, 297, 548


\bibitem[Kondratyuk et al.(1990)]{1990SvAL...16..410K} Kondratyuk, L.~A., Krivoruchenko, M.~I., \& Martemyanov, B.~V.\ 1990, Soviet Astronomy Letters, 16, 410


\bibitem[Kouveliotou et al.(1998)]{1998Natur.393..235K} Kouveliotou, C., Dieters, S., Strohmayer, T., et al.\ 1998, \nat, 393, 235


\bibitem[Kulkarni et al.(2014)]{2014ApJ...797...70K} Kulkarni, S.~R., Ofek, E.~O., Neill, J.~D., Zheng, Z., \& Juric, M.\ 2014, \apj, 797, 70


\bibitem[Kumar et al.(2017)]{2017MNRAS.468.2726K} Kumar, P., Lu, W., \& Bhattacharya, M.\ 2017, \mnras, 468, 2726


\bibitem[Lai \& Shapiro(1991)]{1991ApJ...383..745L} Lai, D., \& Shapiro, S.~L.\ 1991, \apj, 383, 745


\bibitem[Lai et al.(2018)]{2018RAA....18...24L} Lai, X.-Y., Yu, Y.-W., Zhou, E.-P., Li, Y.-Y., \& Xu, R.-X.\ 2018, Research in Astronomy and Astrophysics, 18, 024


\bibitem[Lattimer \& Prakash(2007)]{2007PhR...442..109L} Lattimer, J.~M., \& Prakash, M.\ 2007, \physrep, 442, 109


\bibitem[Li et al.(2017)]{2017RAA....17....6L} Li, L.-B., Huang, Y.-F., Zhang, Z.-B., Li, D., \& Li, B.\ 2017, Research in Astronomy and Astrophysics, 17, 6


\bibitem[Liu(2017)]{2017arXiv171203509L} Liu, X.\ 2017, arXiv:1712.03509


\bibitem[Lor{\'e}n-Aguilar et al.(2009)]{2009A&A...500.1193L} Lor{\'e}n-Aguilar, P., Isern, J., \& Garc{\'{\i}}a-Berro, E.\ 2009, \aap, 500, 1193


\bibitem[Lorimer et al.(2007)]{2007Sci...318..777L} Lorimer, D.~R., Bailes, M., McLaughlin, M.~A., Narkevic, D.~J., \& Crawford, F.\ 2007, Science, 318, 777


\bibitem[Luan \& Goldreich(2014)]{2014ApJ...785L..26L} Luan, J., \& Goldreich, P.\ 2014, \apjl, 785, L26


\bibitem[Lyne \& Lorimer(1994)]{1994Natur.369..127L} Lyne, A.~G., \& Lorimer, D.~R.\ 1994, \nat, 369, 127


\bibitem[Lyubarsky(2008)]{2008ApJ...682.1443L} Lyubarsky, Y.\ 2008, \apj, 682, 1443-1449


\bibitem[Meegan et al.(2009)]{2009ApJ...702..791M} Meegan, C., Lichti, G., Bhat, P.~N., et al.\ 2009, \apj, 702, 791-804


\bibitem[Melrose(1971)]{1971Ap&SS..13...56M} Melrose, D.~B.\ 1971, \apss, 13, 56


\bibitem[Michilli et al.(2018)]{2018Natur.553..182M} Michilli, D., Seymour, A., Hessels, J.~W.~T., et al.\ 2018, \nat, 553, 182


\bibitem[Miralda-Escude et al.(1990)]{1990ApJ...362..572M} Miralda-Escude, J., Paczynski, B., \& Haensel, P.\ 1990, \apj, 362, 572


\bibitem[Nordhaus et al.(2012)]{2012MNRAS.423.1805N} Nordhaus, J., Brandt, T.~D., Burrows, A., \& Almgren, A.\ 2012, \mnras, 423, 1805


\bibitem[Olinto(1987)]{1987PhLB..192...71O} Olinto, A.~V.\ 1987, Physics Letters B, 192, 71


\bibitem[Pagliara et al.(2013)]{2013PhRvD..87j3007P} Pagliara, G., Herzog, M., \& R{\"o}pke, F.~K.\ 2013, \prd, 87, 103007


\bibitem[Palaniswamy et al.(2018)]{2018ApJ...854L..12P} Palaniswamy, D., Li, Y., \& Zhang, B.\ 2018, \apjl, 854, L12


\bibitem[Petroff et al.(2015)]{2015MNRAS.447..246P} Petroff, E., Bailes, M., Barr, E.~D., et al.\ 2015, \mnras, 447, 246


\bibitem[Pizzochero(1991)]{1991PhRvL..66.2425P} Pizzochero, P.~M.\ 1991, Physical Review Letters, 66, 2425


\bibitem[Rees \& M\'{e}sz\'{a}ros(1992)]{1992MNRAS.258P..41R} Rees, M.~J., \& M\'{e}sz\'{a}ros, P.\ 1992, \mnras, 258, 41P


\bibitem[Romero et al.(2016)]{2016PhRvD..93b3001R} Romero, G.~E., del Valle, M.~V., \& Vieyro, F.~L.\ 2016, \prd, 93, 023001


\bibitem[Ruderman \& Sutherland(1975)]{1975ApJ...196...51R} Ruderman, M.~A., \& Sutherland, P.~G.\ 1975, \apj, 196, 51


\bibitem[Scholz et al.(2016)]{2016ApJ...833..177S} Scholz, P., Spitler, L.~G., Hessels, J.~W.~T., et al.\ 2016, \apj, 833, 177


\bibitem[Shibata \& Magara(2011)]{2011LRSP....8....6S} Shibata, K., \& Magara, T.\ 2011, Living Reviews in Solar Physics, 8, 6


\bibitem[Spitler et al.(2016)]{2016Natur.531..202S} Spitler, L.~G., Scholz, P., Hessels, J.~W.~T., et al.\ 2016, \nat, 531, 202


\bibitem[Stejner \& Madsen(2005)]{2005PhRvD..72l3005S} Stejner, M., \& Madsen, J.\ 2005, \prd, 72, 123005


\bibitem[Thompson \& Duncan(1995)]{1995MNRAS.275..255T} Thompson, C., \& Duncan, R.~C.\ 1995, \mnras, 275, 255


\bibitem[Thornton et al.(2013)]{2013Sci...341...53T} Thornton, D., Stappers, B., Bailes, M., et al.\ 2013, Science, 341, 53


\bibitem[Timmes et al.(1996)]{1996ApJ...457..834T} Timmes, F.~X., Woosley, S.~E., \& Weaver, T.~A.\ 1996, \apj, 457, 834


\bibitem[Totani(2013)]{2013PASJ...65L..12T} Totani, T.\ 2013, \pasj, 65, L12


\bibitem[Usov(1997)]{1997ApJ...481L.107U} Usov, V.~V.\ 1997, \apjl, 481, L107


\bibitem[Usov(1998a)]{1998PhRvL..81.4775U} Usov, V.~V.\ 1998a, Physical Review Letters, 81, 4775


\bibitem[Usov(1998b)]{1998PhRvL..80..230U} Usov, V.~V.\ 1998b, Physical Review Letters, 80, 230


\bibitem[Usov(2001a)]{2001ApJ...550L.179U} Usov, V.~V.\ 2001a, \apjl, 550, L179


\bibitem[Usov(2001b)]{2001PhRvL..87b1101U} Usov, V.~V.\ 2001b, Physical Review Letters, 87, 021101



\bibitem[Wang et al.(2018)]{2018ApJ...852..140W} Wang, W., Luo, R., Yue, H., et al.\ 2018, \apj, 852, 140


\bibitem[Witten(1984)]{1984PhRvD..30..272W} Witten, E.\ 1984, \prd, 30, 272


\bibitem[Xu \& Busse(2001)]{2001A&A...371..963X} Xu, R.~X., \& Busse, F.~H.\ 2001, \aap, 371, 963


\bibitem[Xu et al.(2001)]{2001APh....15..101X} Xu, R.~X., Zhang, B., \& Qiao, G.~J.\ 2001, Astroparticle Physics, 15, 101


\bibitem[Zhang(2014)]{2014ApJ...780L..21Z} Zhang, B.\ 2014, \apjl, 780, L21


\end{thebibliography}
\end{document}